\documentclass[prb,aps,twocolumn,superscriptaddress]{revtex4-1}
\usepackage{amsmath}
\usepackage{latexsym}
\usepackage[english]{babel}
\usepackage{graphicx}
\DeclareGraphicsExtensions{.pdf,.png,.jpg, .jpeg, .eps}
\usepackage{bm}
\usepackage{array}
\usepackage{amssymb}
\usepackage{amsfonts}
\usepackage{multirow}
\usepackage{epstopdf}
\usepackage{color}

\begin{document}
	
	\title{Phase transitions in superconductor/ferromagnet bilayer driven by spontaneous supercurrents }

	\author{Zh. Devizorova}
	\affiliation{Moscow Institute of Physics and Technology, 141700 Dolgoprudny, Russia}
	\author{A. V. Putilov}
	\affiliation{Institute for Physics of Microstructures, Russian Academy of Sciences, 603950 Nizhny Novgorod, GSP-105, Russia}
	\author{I. Chaykin}
	\affiliation{Moscow Institute of Physics and Technology, 141700 Dolgoprudny, Russia}
	\affiliation{Kotelnikov Institute of Radioengineering and Electronics of Russian Academy of Sciences, Moscow, 125009, Russia}
	\author{S. Mironov}
	\affiliation{Institute for Physics of Microstructures, Russian Academy of Sciences, 603950 Nizhny Novgorod, GSP-105, Russia}
	\author{A.I. Buzdin}
	\affiliation{University Bordeaux, LOMA UMR-CNRS 5798, F-33405 Talence Cedex, France}
	\affiliation{World-Class Research Center “Digital biodesign and personalized healthcare”, Sechenov First Moscow State Medical University, Moscow 119991, Russia}
	
	\begin{abstract}
		We investigate superconducting phase transition in superconductor(S)/ferromagnet(F) bilayer with Rasba spin-orbit interaction at S/F interface. This spin-orbit coupling produces spontaneous supercurrents flowing inside the atomic-thickness region near the interface, which are compensated by the screening Meissner currents [S. Mironov and A. Buzdin, Phys. Rev. Lett \textbf{118}, 077001 (2017)]. In the case of thin superconducting film the emergence of the spontaneous surface currents causes the increase of the superconducting critical temperature and we calculate the actual value of the critical temperature shift. We also show that in the case of type-I superconducting film this phase transition can be of the first order. In the external magnetic field the critical temperature depends on the relative orientation of the external magnetic field and the exchange field in the ferromagnet. Also we predict the in-plane anisotropy of the critical current which may open an alternative way for the experimental observation of the spontaneous supercurrents generated by the SOC.
	\end{abstract}
	
	\maketitle
	\section{Introduction}
The superconducting states carrying spontaneous current in the systems with broken time reversal symmetry were the subject of interest for more than twenty years \cite{Fogelstrom, Walter, Matsumoto, Barash, Honerkamp, Kwon,  Vorontsov, Hakansson, Bobkova, Roy}. Such spontaneous supercurrents were predicted for $d$-wave \cite{Fogelstrom, Walter, Barash, Honerkamp} or chiral $p$-wave \cite{Matsumoto, Roy} superconductors, for a mesoscopic normal metal film in contact with a superconductor \cite{Blatter} and at the interface between a superconductor and a ferromagnet \cite{Bobkova}. These currents are typically carried by Andreev edges states \cite{Fogelstrom, Walter, Matsumoto, Barash, Honerkamp, Kwon,  Vorontsov, Hakansson, Bobkova, Roy} and appear at temperature $T$ well below the superconducting critical temperature $T_c$. 

Recently the spontaneous supercurrents were predicted to appear at the interface between s-wave superconductor (S) and a ferromagnetic (F) insulator \cite{Mironov}. Contrary to the spontaneous supercurrents carried by Andreev bound states \cite{Fogelstrom, Walter, Matsumoto, Barash, Honerkamp, Kwon,  Vorontsov, Hakansson, Bobkova, Roy}, these currents appears at the superconducting transition, i.e. at $T=T_c$. The crucial condition for the emergence of these currents is the presence of Rashba \cite{Bychkov} spin-orbit coupling  (SOC) at the S/F interface \cite{Mironov}. Indeed, this SOC produces the additional term $\propto (\sigma \times {\bf p}) \cdot {\bf n}$ in the effective Hamiltonian of a conducting electron (${\bf n}$ is the unit vector perpendicular to the S/F interface). As a result, spin and momentum appear to be coupled, which produces the nontrivial “helicity” of the electronic energy bands. Since the exchange field makes the spin-up state energetically more favourable than the spin-down one, one may expect the emergence of the electric current. Note that the  helical states \cite{Edelstein, Mineev} also play an important role in the emergence of Majorana modes \cite{Alicea}, the formation of Josephson $\varphi_0$ junctions with spontaneous nonzero phase difference across the junction in the ground state \cite{Buzdin,Mironov2, Robinson, Krive,Reynoso, Szombati}, and the appearance of Fulde-Ferrell-Larkin-Ovchinnikov (FFLO) like states with finite Cooper-pair momentum \cite{Smidman}.

Since spontaneous supercurrents flowing at the S/F interface with SOC appear at $T=T_c$, these  currents can affect the parameters of the phase transition (the superconducting critical temperature, the phase diagram in the external magnetic field, the critical current, etc). Although the spontaneous supercurrents result in the local enhancement of the superconductivity near the S/F interface, it was shown \cite{Mironov} that for the large thickness of the superconductor they do not affect the superconducting transition temperature and, thus, with the increase in temperature the superconductivity becomes destroyed in the whole bulk of the sample. This situation is in contrast with the well-known phenomena arising in the superconductors containing twinning planes which locally increase the critical temperature and favor the emergence of the localized superconducting states above the bulk critical temperature (see, e.g., Ref.~\onlinecite{Buzdin_TP} for review). 

In the present paper we study the effect of the spontaneous supercurrents on the superconducting phase transition in S/F bilayer with finite thickness of the superconducting layer and SOC of the Rashba type at the S/F interface. We show that in the case of thin superconducting film these currents causes the increase of the superconducting critical temperature $T_c$ and we calculate the corresponding critical temperature shift. Surprisingly, in type-I superconductors the superconducting phase transition is of the first order even in the absence of external magnetic field. At the same time, in external magnetic field the critical temperature strongly depends on the relative orientation between external magnetic field and the exchange field in the ferromagnet. Note that in the case of positive (negative) SOC parameter $T_c$ is significantly higher (lower) for the parallel magnetic configuration in comparison with the antiparallel one. At the same time, in the case of type-II superconducting layer the emergence of the spontaneous supercurrents also increases the superconducting critical temperature in the absence of external magnetic field and makes in sensitive to the relative orientation of external magnetic field and the exchange field, if the sample placed into the field, but the phase transition is of the second order. All described phenomena can serve the hallmarks of the spontaneous supercurrents and can be used for the experimental detection of these currents. 

The paper is organized as follows. In section \ref{Section1} we introduce the model and study the phase transition in the S/F bilayer with type-I superconductor. In section \ref{Section2} we calculate the dependence of the critical temperature on the external magnetic field. In section \ref{Section3} we consider type-II superconductor. In section \ref{Section4} we analyze the anisotropy of the critical current. In section \ref{Concl} we summarize our results.

\section{First order phase transition in S/F bilayer with thin type-I superconductor}
\label{Section1}
\begin{figure}[b!]
		\includegraphics[width=0.7\linewidth]{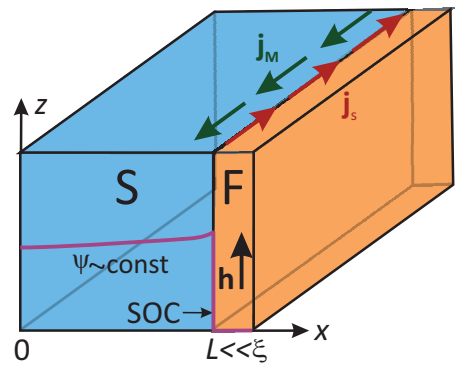}
		\caption{The sketch of a superconducting (S) film placed in contact with thin ferromagnetic (F) layer. The spin-orbit coupling at the S/F inteface produces spontaneous supercurrent $j_s$ causing the increase of the superconducting order parameter inside the superconductor.} \label{Fig:sample}
	\end{figure}

We consider thin superconducting (S) film of the thickness $L$ placed in contact with a ferromagnetic insulator (F), see Fig.\ref{Fig:sample}. We assume the presence of Rashba spin-orbit coupling (SOC) at the S/F interface. The free energy functional of the system under consideration is given by the following expression\cite{Mironov, Samokhin, Kaur}:
\begin{multline}
F = \iiint dV \biggl\{\alpha |\psi|^2 + \frac{\beta}{2}|\psi|^4 + \frac{1}{4m}|\hat{\boldsymbol{D}}\psi|^2 + \\ \frac{(\rm{rot} \boldsymbol{A})^2}{8\pi} + [\boldsymbol{n}\times \boldsymbol{h}]\epsilon(r)\left(\psi^*\hat{\boldsymbol{D}}\psi + \psi\hat{\boldsymbol{D^\dagger}}\psi^*\right)\biggr\}.
\end{multline}
Here $\alpha = a(T-T_{c0})$ and $\beta$ are the standard GL coefficients, $T_{c0}$ is the critical temperature of bulk superconductor, $\hat{\boldsymbol{D}} = -i\hbar\boldsymbol{\nabla} + \frac{2e}{c}\boldsymbol{A}$, ${\bf A}$ is the vector-potential, $\boldsymbol{n}$ is the unit vector perpendicular to the S/F interface and directed from S to F layer, $\boldsymbol{h}$ is the exchange field in the ferromagnet, $\psi$ is the order parameter of the superconductor and $\epsilon$ is the spin-orbit constant, which is nonzero only in the atomically thin area of the width $l_{so}$ near the S/F interface. Without loss of generality, let us assume that the exchange field is directed along the 
$z$-axis, i.e. $\boldsymbol{h}=h {\bf e}_z$, the $x$-axis coincides with ${\bf n}$ and the S/F interface is located at $x=L$.

In such system spontaneous supercurrents flow near the S/F interface \cite{Mironov}. These currents cause the increase of the superconducting order parameter in the area of the width $\sim \xi$ (superconducting coherence length) near S/F interface. If the thickness $L$ of the superconducting film is much smaller than $\xi$, i.e. $L \ll \xi$, one can expect the increase of the superconducting critical temperature $T_c$. Here we find  the actual temperature shift for type-I superconductor, i.e. for the case $\lambda \ll L \ll \xi$, where $\lambda$ is the London penetration depth. Also we show, that the phase transition is of the first order. For this purpose, let us calculate the free energy for such a system. Due to the condition $L \ll \xi$ we may take $\psi(x) \approx const$. Let us introduce the dimensionless order parameter $\varphi = \sqrt{\beta/|\alpha_0|}\psi$, where $\alpha_0 = -aT_{c0}$. 

Since the spontaneous supercurrents are fully compensated by the Meissner currents,  the magnetic field ${\bf B}=(0,0,B_z)$ is absent outside the superconductor and has the following form: 
	
	\begin{equation}
	\label{Ay}
	B_z(x) = \left[ \begin{aligned}
    & B_0{\rm exp}\left(\frac{x-L}{\lambda}\right),\,\,\ 0<x<L , \\ 
    & 0,\,\,  \qquad x>L, \qquad x\le 0. \\ 
\end{aligned} \right.
	\end{equation}
Here $\lambda=\lambda_0/\varphi$, where  $\lambda_0^2 = mc^2\beta/(8\pi e^2|\alpha_0|)$ is the zero-temperature London penetration depth for the bulk superconductor. Note that inside the superconducting slab we neglect the second exponent [see the first line in Eq.(\ref{Ay})] since we assume $\lambda \ll L$. The resulting magnetic field Eq.(\ref{Ay}) is continuous at $x=0$ and experiences a jump by the value $B_0$ at $x=L$ due to the surface spontaneous supercurrents flowing along S/F interface. To find the actual value of the jump we substitute the magnetic field to the free energy and minimize it with respect to $B_0$. We obtain:
	
	\begin{equation}
	A_0 = \Delta H\varphi^2.
	\end{equation}
	Here $\Delta H=4\sqrt{2}H_{cm}k_{so} l_{so} (\xi_0/\lambda_0)$ is the jump of magnetic field at S/F interface due to spontaneous surface supercurrents, where $H_{cm}=\sqrt{4\pi\alpha_0^2/\beta}$ is the thermodynamic critical magnetic field, $k_{so}=m h \epsilon/\hbar$ and $\xi_0=\sqrt{\hbar^2/4m|\alpha_0|}$.
	
The resulting free energy reads as:
	
	\begin{equation}
	F = V\biggl(\frac{\alpha|\alpha_0|}{\beta}\varphi^2+\frac{|\alpha_0|^2}{2\beta}\varphi^4\biggr) - S\frac{\Delta H^2}{8\pi}\lambda_0\varphi^3,
	\end{equation}
where $V$ is the volume of the superconducting slab and $S$ is the surface area of S/F boundary.
	
We find the value of the critical temperature $T_c$ and the order parameter $\varphi_{cr}$ at $T=T_c$ by minimizing $F$ with respect to $\varphi$ and using the condition $F=0$, which is fulfilled at the critical temperature. We find:
	\begin{equation}
	\varphi_{cr} = \frac{1}{2}\frac{\lambda_0}{L}\biggl(\frac{\Delta H}{H_{cm}}\biggr)^2,
	\end{equation}
	\begin{equation}
	\label{T_c1}
	\frac{T_c}{T_{c0}} = 1 + \frac{1}{8}\biggl(\frac{\lambda_0}{L}\biggr)^2\biggl(\frac{\Delta H}{H_{cm}}\biggr)^4.
	\end{equation}
	
One see that the critical temperature, indeed, increases due to the spin-orbit interaction at the S/F interface. Moreover, since the order parameter does not equal to zero at the transition temperature, in the structure under consideration the phase transition is of the first order. Note that the critical temperature Eq.(\ref{T_c1}) is not divergent when the slab thickness $L$ tends to zero, since we consider the situation $\lambda \ll L$.

The obtained result is valid only if the assumption $\lambda \ll L$ is fulfilled at $T=T_c$. To check this, we find the London penetration depth at the critical temperature:
	
	\begin{equation}
	\lambda_{cr} = \frac{2L}{\bigl(\frac{\Delta H}{H_{cm}}\bigr)^2}.
	\end{equation}
	Since $\Delta H/H_{cm} \sim k_{so}l_{so} (\xi_0/\lambda_0)$ and $(\xi_0/\lambda_0) \gg 1$ the assumption $\lambda_{cr} \ll L$ is valid if $k_{so}l_{so}$ is not very small.

	Let us estimate $\Delta T_c=T_c-T_{c0}$. Since $\epsilon =v_{so}/E_F$, where $v_{so}$ is the spin-orbit velocity and $E_F$ is the Fermi energy, the jump of the magnetic field at $x=L$ due to the spontaneous supercurrents can be estimated as $\Delta H \approx 4\sqrt{2}H_{cm}(h/E_F)(v_{so}/v_F)(\xi_0/\lambda_0)$, where $v_F$ is the Fermi velocity. It is reasonable to take $h/E_F \sim 0.1$, $v_{so}/v_F \sim 0.1$. If we also assume $\xi_0/\lambda_0 \sim 50$, then we find $\Delta H \approx 3H_{cm}$. Taking $\lambda_0/L \sim 0.1$ we obtain $\Delta T_c/T_{c0} \approx 0.1$. Since $\Delta T \ll T_{c0}$ the Ginzburg-Landau approach is applicable.
	
Note that the first-order phase transition in S/F bilayer with Rashba-type spin-orbit interaction and $\lambda_0 \ll L \ll \xi$ can serve as a hallmark of the spontaneous supercurrents flowing at S/F interface \cite{Mironov}.

The suitable system for the observation of the discussed effects may be based on thin epitaxially grown layers of extreme type-I superconductors. For example, recently it was demonstrated the epitaxial growth of high-quality single-crystalline aluminum films \cite{Cheng}.
	
\section{Phase transition in external magnetic field}
\label{Section2}
In the external magnetic field ${\bf H}_0$ the critical temperature of superconducting phase transition depends on the relative orientation of ${\bf H}_0$ and the exchange field ${\bf h}$. For simplicity, let us restrict ourselves to the case ${\bf H}_0=H_0 {\bf e}_z$, where $H_0$ can be both positive and negative. To find actual $T_c(H_0)$ dependence let us write down the Gibbs free energy of the system:
	
		\begin{multline}
		\label{Eq:G}
		G = \iiint dV \biggl\{\alpha |\psi|^2 + \frac{\beta}{2}|\psi|^4 + \frac{1}{4m}|\hat{\boldsymbol{D}}\psi|^2 + \\ \frac{(\bm{B} - \bm{H_0})^2}{8\pi} +  [\boldsymbol{n}\times \boldsymbol{h}]\epsilon(r)\left(\psi^*\hat{\boldsymbol{D}}\psi + \psi\hat{\boldsymbol{D^\dagger}}\psi^*\right)\biggr\}.
		\end{multline}
		
 As before, we consider the case $L \ll \xi$ and assume $\psi \approx {\rm const}$. Since we also assume $\lambda \ll L$, the magnetic field ${\bf B} =B \bm{e}_z= {\rm rot} \bm{A}$ inside the superconducting slab reads:
		
		\begin{equation}
		B = H_0{\rm exp}\left(-\frac{x}{\lambda}\right) + \left(H_0+\Delta H\varphi^2\right){\rm exp}\left(\frac{x-L}{\lambda}\right).
		\end{equation}

 Calculating the resulting Gibbs free energy we obtain:	
	
	\begin{multline}
	\label{Eq_G}
	\frac{G}{V} = \frac{\alpha|\alpha_0|}{\beta}\varphi^2 + \frac{|\alpha_0|^2}{2\beta}\varphi^4 +\frac{H_0^2}{8\pi} - \frac{H_0\Delta H}{4\pi}\frac{\lambda_0}{L}\varphi - \frac{\Delta H^2}{8\pi}\frac{\lambda_0}{L}\varphi^3
	\end{multline}
Note that the surface contribution to the free energy caused by spontaneous supercurrents results in the increases	(decreases) of the free energy if ${\bf H}_0 \uparrow \uparrow {\bf h}$ (${\bf H}_0 \uparrow \downarrow {\bf h}$).  
	
The critical temperature and the order parameter at the critical point $\varphi_{cr}$ can be found from the system of equations $G=0$ and $\partial_{\varphi} G=0$, which are fulfilled at $T=T_c$. From the second equation we find the critical temperature as a function of $\varphi_{cr}$:
	
	\begin{multline}
	\label{Tc_H0}
	\frac{T_c}{T_{c0}}=1-\varphi_{cr}^2 +\frac{3}{4}\left(\frac{\Delta H}{H_{cm}}\right)^2\frac{\lambda_0}{L}\varphi_{cr} +\\+\frac{1}{2\varphi}\frac{H_0}{H_{cm}}\frac{\Delta H}{H_{cm}}\frac{\lambda_0}{L}.
	\end{multline}
	
At the same time, the order parameter obeys the equation
	\begin{multline}
	\label{phi_crit}
	\varphi_{cr}^4-\frac{1}{2}\left(\frac{\Delta H}{H_{cm}}	\right)^2 \frac{\lambda_0}{L} \varphi_{cr}^3 -\left(\frac{H_0}{H_{cm}}\right)^2 +\\+\frac{H_0}{H_{cm}}\frac{\Delta H}{H_{cm}}\frac{\lambda_0}{L}\varphi_{cr}=0.
	\end{multline}
	
\begin{figure}[t!]
		\includegraphics[width=0.7\linewidth]{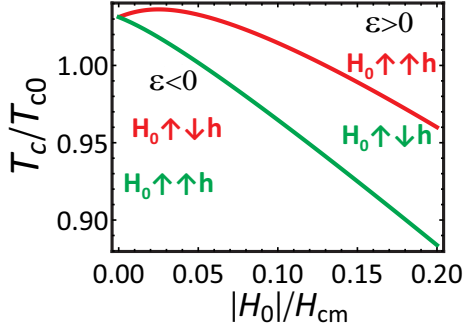}
		\caption{The dependence of the superconducting critical temperature $T_c$ of the S/F bilayer with $\lambda_0=0.02 L$ and $\Delta H=5 H_{cm}$ on the external magnetic field ${\bf H_0}=H_0 {\bf e}_z$. Here $\epsilon$ is the spin-orbit coupling constant. }\label{Fig:Tc_H0}
	\end{figure}

Solving the equations we indeed find that the critical temperature of the superconducting phase transition strongly depends on the relative orientation of the external magnetic field and the exchange field: for positive (negative) SOC parameter $\epsilon$ the $T_c$ is higher for the parallel (anti-parallel) orientation compared to the anti-parallel (parallel) one (see Fig. \ref{Fig:Tc_H0}). This findings provides a tool for the experimental observation of the predicted effect, although the correction to $T_c$ due to SOC is small. Changing the direction of the external magnetic field, one can observe the variation of superconducting critical temperature. Note that the dependence of the critical temperature on the magnetic configuration is also can be used for the experimental detection of the sign  of the spin-orbit parameter. Indeed, the critical temperature is higher for the parallel orientation between the external magnetic field and the exchange field in comparison with the antiparallel one only if the spin orbit constant is positive, while for the negative SOC parameter the situation is reversed. 
	
Since $\lambda_0 \ll L$ we can find approximate analytical expressions for $T_c^{\uparrow \uparrow}$ and  $T_c^{\uparrow \downarrow}$ expanding the results obtained from Eq.(\ref{Tc_H0}) and Eq.(\ref{phi_crit}) over $\lambda_0/L$. To have good agreement between the exact results and the approximate one, we should expand these expressions up to the third order over $\lambda_0/L$. If $\Delta H>0$ ($\Delta H<0$) for the parallel (anti-parallel) magnetic configuration we obtain:

\begin{multline}
    \frac{T_c^{\uparrow \uparrow (\uparrow \downarrow)}}{T_{c0}}=1-\frac{|H_0|}{H_{cm}}+\frac{1}{2}\sqrt{\frac{|H_0|}{H_{cm}}} \frac{\Delta H}{H_{cm}}\left(2+\frac{\Delta H}{H_{cm}}\right)\frac{\lambda_0}{L} +\\+\frac{1}{32}\left(\frac{\Delta H}{H_{cm}}\right)^2  \left[\left(\frac{\Delta H}{H_{cm}}\right)^2-2 \right]^2 \left(\frac{\lambda_0}{L} \right)^2.
\end{multline}

At the same time, if $\Delta H>0$ ($\Delta H<0$) for the antiparallel (parallel) magnetic configuration we find:
\begin{multline}
    \frac{T_c^{\uparrow \downarrow (\uparrow \uparrow )}}{T_{c0}}=1-\frac{|H_0|}{H_{cm}}-\sqrt{\frac{|H_0|}{H_{cm}}} \frac{\Delta H}{H_{cm}}\frac{\lambda_0}{L} +\\+\frac{1}{\sqrt{3}}\left(\frac{\Delta H}{H_{cm}}\right)^2  \left(0.9\frac{H_0}{H_{cm}}+\frac{\sqrt{3}}{8}\right)\left(\frac{\lambda_0}{L} \right)^2.
\end{multline}

Note that the above expressions are invalid, when $H_0$ tends to zero. Assuming $H_0 \ll \Delta H$ from Eq.(\ref{Tc_H0}) and Eq.(\ref{phi_crit}) we find:

	\begin{equation}
	    \frac{T_c^{\uparrow \uparrow(\uparrow \downarrow)}}{T_{c0}}=1+\frac{1}{8}\left(\frac{\Delta H}{H_{cm}}\right)^4\left(\frac{\lambda_0}{L}\right)^2 \pm \frac{2|H_0|}{\Delta H}.
	\end{equation}

Note that the dependence of the critical temperature on the in plane magnetic field orientation is another hallmark of spontaneous supercurrents flowing at the S/F interface \cite{Mironov} (in addition to the appearance of the stray magnetic field near S/F interface and the anisotropy of the upper critical field predicted in Ref.[\onlinecite{Mironov}]). This fact can be used for the experimental detection of the spontaneous supercurrents.

\section{Phase transition in the case of thin type-II superconducting layer}
\label{Section3}

In this section we analyze the peculiarities of the superconducting phase transition in the S/F bilayer (see Fig.~\ref{Fig:sample}) for the case when the superconducting layer is of the II type. We show that in this case the critical temperature also increases in the absence of the external magnetic field ${\bf H}_0$ and depends on the relative orientation between the external magnetic field and the exchange field. Let us consider the temperatures close to the superconducting transition temperature and the external magnetic field directed along the $z$ axis so that ${\bf H}_0=(0,0,H_0)$. At the point of the phase transition the superconducting order parameter is small which allow us to neglect the term $\propto \psi |\psi|^2$ and the term associated with the spontaneous supercurrents in the Ginzburg-Landau (GL) equation. Choosing the gauge of the vector potential in the form ${\bf A}=(0,H_0x,0)$ and searching the solution in the form $\psi =e^{ik_y y}\psi(x)$ we obtain the following GL equation:
\begin{equation}
    \alpha \psi(x)-\frac{\hbar^2}{4m}\partial_{xx}\psi(x)+\frac{\left(\hbar k_y+2eH_0x/c\right)}{4m}\psi(x)=0,
\end{equation}
with the boundary conditions
\begin{equation}
    \partial_x\psi(0)=0, \qquad \partial_x \psi(L)=\frac{8m h \epsilon l_{so}}{\hbar^2}\left(\hbar k_y+\frac{2eH_0L}{c}\right)\psi(L).
\end{equation}

In the absence of the external magnetic field the order parameter reads as $\psi(x)=A{\rm cosh}qx$, where $q^2=\left(4m\alpha/\hbar^2 +k_y^2\right)$. Calculating the free energy, we find:
\begin{multline}
    \frac{F}{S}=\frac{A^2L}{2}\left(\alpha+\frac{\hbar^2k_y^2}{4m}\right)\left(\frac{\sinh{2ql}}{2qL}+1\right)+A^2\frac{\hbar^2q^2}{4m}\times\\ \times \frac{L}{2} \left(\frac{\sinh{2ql}}{2qL}-1\right)-2A^2 h\epsilon l_{so}\hbar k_y\cosh^2qL.
\end{multline}

Assuming $qL\ll 1$ and minimizing the free energy with respect to $k_y$, we find the optimal modulation vector $k_y=4k_{so}l_{so}/L$. Since at the critical point F=0, we obtain the increase of the superconducting critical temperature:
\begin{equation}
    \frac{T_c-T_{c0}}{T_{c0}}=16k_{so}^2l_{so}^2\left(\frac{\xi_0}{L}\right)^2.
 \end{equation}
Note that the effect is absent in the limit $L \rightarrow \infty$, i.e. for thick superconducting slab.   
    
When the sample is placed in the external magnetic field it is convenient to choose the origin of the $x$-axis at the center of S film, so the S/F boundary is located at $x=L/2$ and the other boundary at $x=-L/2$. Following the procedure described in Ref.~[\onlinecite{James}] we introduce the dimensionless coordinate $X=2x/L$, the modulation vector $K_y=k_yL/2$, the parameters $\tilde{H}_0=eH_0L^2/(2\hbar c)$, $\epsilon_0=-m\alpha L^2/\hbar^2$  and rewrite the GL equation in the following form:
\begin{equation}
    \partial_{XX}\psi(X)+(K_y+\tilde{H}_0X)^2\psi(X)=\epsilon_0 \psi(X).
\end{equation}
At the same time, the boundary conditions read
\begin{equation}
    \partial_X \psi(1)=s(K_y+\tilde{H}_0)\psi(1), \qquad \partial_{X} \psi(-1)=0,
\end{equation}
where $s=8k_{so}l_{so}$. 

Next it is useful to introduce the new variable $t=\sqrt{2|\tilde{H}_0|}(X+K_y/h_0)$. The resulting GL equation and the boundary conditions are  the following:
\begin{equation}
\label{Weber}
     -\partial_{tt}\psi(t)+\frac{1}{4}t^2\psi(t)=\frac{\epsilon_0}{2|\tilde{H}_0|}\psi(t),
\end{equation}
\begin{gather}
        \partial_t \psi|_{\sqrt{2|\tilde{H}_0|}(-1+K_y/\tilde{H}_0)}=0,\\
       \left.\left(\frac{\partial_t \psi}{\psi}\right)\right|_{\sqrt{2|\tilde{H}_0|}(1+K_y/\tilde{H}_0)}=\frac{s(K_y+\tilde{H}_0)}{\sqrt{2|\tilde{H}_0|}}.
\end{gather}
   
The solution of Eq.(\ref{Weber}) has the form of the linear combination of the Weber functions $\psi(t)=A_{\nu}D_{\nu}(t)+B_{\nu}D_{\nu}(-t)$, where $2\nu+1=\epsilon_0/|\tilde{H}_0|$. Substituting it into the boundary conditions we obtain the following equation, which implicitly defines the function $T_c(H_0)$:
\begin{multline}
\label{nu}
    D'_{\nu}(\alpha_+) D'_{\nu}(-\alpha_-)-D'_{\nu}(-\alpha_+) D'_{\nu}(\alpha_-)=\frac{s(K_y+\tilde{H}_0)}{\sqrt{2|\tilde{H}_0|}}\\\times \biggl[D_{\nu}(\alpha_+) D'_{\nu}(-\alpha_-)-D_{\nu}(-\alpha_+) D'_{\nu}(\alpha_-)\biggr],
\end{multline}
where $\alpha_{\pm}=\sqrt{2|\tilde{H}_0|}(\pm 1 +K_y/\tilde{H}_0)$.

\begin{figure}[t!]
		\includegraphics[width=0.7\linewidth]{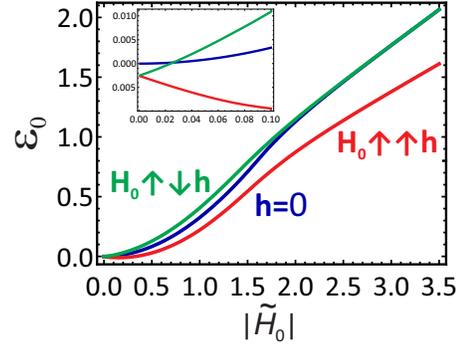}
		\caption{The dependence of the critical temperature $T_c$ on the external magnetic field ${\bf H}_0=(0,0,H_0)$. Here  $\epsilon_0=\left(1-T_c/T_{c0}\right)L^2/(4\xi_0^2)$ and $\tilde{H}_0=eH_0L^2/(2\hbar c)$.}\label{Fig:eps_h0}
	\end{figure}

The maximal value of $T_c$ at fixed magnetic field corresponds to the minimal value of $\nu$, which satisfies Eq.(\ref{nu}). At fixed $K_y$ and $\tilde{H}_0$ this equation has infinite but discrete number of solutions for $\nu$, and we find the minimal one. Then we minimize it with respect to $K_y$ and find the minimal value $\nu_0$ for fixed $\tilde{H}_0$. As a result, we obtain the dependence $\epsilon_0(\tilde{H}_0)$ (i.e. $T_c(H_0)$) in the form $\epsilon_0=(2\nu_0+1)|\tilde{H}_0|$, see. Fig.\ref{Fig:eps_h0}. The critical temperature is higher for the parallel orientation of the external magnetic field and the exchange field in comparison with the antiparallel one. Here we assume that the spin-orbit coupling parameter $\epsilon$ is positive.

As we can see from the inset in Fig.\ref{Fig:eps_h0} the weak parallel magnetic field leads to the initial increase of the critical temperature, which is replaced by the usual decrease at higher magnetic field. Such peculiar behavior remind the experimentally observed in Ref.[\onlinecite{Gardner}] increase of the critical temperature in thin Pb films. We may speculate that the local  SOC could be generated at Pb/substrate  interface, while the role of the exchange field is played by a Zeeman field.

	\section{Critical current}
	\label{Section4}
	
In this section we show that the spin-orbit coupling makes the in-plane critical current of S/F bilayer anisotropic. Although the total spontaneous superconducting current generated by the SOC is zero, it is non-uniformly distributed across the layers. As a result, for the fixed direction of the exchange field the local current density (and, thus, the local damping of the superconducting order parameter) at a certain point of the S film becomes dependent on the angle $\theta$ between the external current and the spontaneous current flowing along the S/F interface. Consequently, 
the maximal current which does not destroy the superconducting state (critical current) also becomes dependent on $\theta$ (diode-like effect).


To calculate the critical current of the S/F bilayer we again consider the system sketched in Fig.~\ref{Fig:sample}. First, we consider the situation when the external transport current of the linear density $J$ is directed along the $y$ axis. Since the magnetic field produced by both the current $J$ and the spontaneous surface current due to SOC is directed along the $z$ axis and depends only on the $x$ axis we may choose the vector potential in the form ${\bf A}=A(x)\hat{\bf y}_0$. Also we choose the order parameter $\varphi$ to be real. Then the density of the free energy accounting the non-uniform profile of the order parameter $\varphi$ and the vector potential $A$ across the structure can be written in the form
    \begin{multline}\label{FE_nonuniform}
		\frac{F}{V} = \frac{H_{cm}^2}{4\pi}\left(
		-\tau\varphi^2+\varphi^4+\xi_0^2\varphi'^2
		\right)+\\
		+\frac{A'^2}{8\pi}+\frac{A^2\varphi^2}{8\pi\lambda_0^2}-\frac{\Delta HA\varphi^2}{4\pi}\delta(x-L),
	\end{multline}
where $\tau=1-T/T_{c0}$, $\varphi^\prime\equiv \partial\varphi/\partial x$ and $A^\prime\equiv \partial A/\partial x$. Varying the free energy with respect to $\varphi(x)$ and $A(x)$ inside the S layer we drive to the standard set of the Ginzburg-Landau equations 
\begin{align}
    -\xi_0^2\varphi''-\tau\varphi+\varphi^3+A^2\varphi/(2H_{cm}^2\lambda_0^2)&=0,
    \label{Eq_GL}\\
    -\lambda_0^2 A''+\varphi^2 A&=0,
    \label{Eq_L}
\end{align}
supplemented by the boundary conditions accounting the surface energy contribution due to the spin-orbit coupling [the last term with $\delta$--function in Eq.~(\ref{FE_nonuniform})] and the magnetic field $\pm (2\pi/c)J$ generated by the external transport current $J$ at the outer boundaries $x=0$ and $x=L$ of the superconducting film:
\begin{align}
    & \varphi'(0)=0,\,\,\varphi'(L)=\Delta HA(L)\varphi(L)/H_{cm}^2,
    \label{Eq_BC_Phi}\\
    & A'(0)=-(2\pi/c)J,\,\,\,A'(L)=(2\pi/c)J+\Delta H\varphi^2(L).
    \label{Eq_BC_A}
\end{align}

The accurate solution of Eqs.~(\ref{Eq_GL})-(\ref{Eq_L}) requires focusing on two features responsible for the anisotropy of the critical current. The first one is the non-uniform distribution of the screening Meissner current across the S film. The second one is the damping of the order parameter at the S/F interface by the transport current and the subsequent renormalization of the spontaneous surface current (and the screening Meissner one). Thus, although in the case of thin S layer the terms containing spatial derivatives of the order parameter and the vector potential cannot be neglected.

In order to find the analytical solution of the Ginzdurg-Landau equations we make several assumptions simplifying the calculations. First, we restrict ourselves to the most interesting case of the type-II superconductor and assume that the thickness $L$ of the S film is much smaller than the superconducting coherence length so that $L\ll\xi\ll\lambda$. Second, we consider the limit of small spin-orbit coupling assuming the dimensionless parameter $\mu=\Delta H\lambda_0/(H_{cm}L)$ to be small ($\mu\ll 1$). These assumptions allow us to expand the functions $A(x)$ and $\varphi(x)$ over $x$ keeping the terms up to $(L/\xi)^3$ in order to account the non-uniform distribution of the superconducting current across the S film:
\begin{align*}
    & A=A_0+A_1x+A_2x^2+A_3x^3,\\
    & \varphi=\varphi_0+\varphi_1x+\varphi_2x^2+\varphi_3x^3.    
\end{align*}
Also in the resulting perturbation theory it is enough to consider the terms proportional to $\mu$ and neglect the higher order contributions.

Substituting the expansion for $A$ and $\varphi$ into the equations (\ref{Eq_GL})--(\ref{Eq_L}) and the boundary conditions (\ref{Eq_BC_Phi})--(\ref{Eq_BC_A}), we find the dependence of the external current $J$ on the dimensionless vector potential $a_0=A_0/(H_{cm}\lambda_0)$:
\begin{multline}
    J=cH_{cm}L/(8\pi\lambda_0)
    \left(
    2\tau a_0-2\tau\mu+3\mu a_0^2-a_0^3
    \right)\times\\
    \times\left[1-L^2(2\tau+4a_0\mu-3a_0^2)/(8\lambda_0^2)
\right]
\end{multline}
Then the critical current $J_c$ of the S/F bilayer can be obtained as the maximum of the dependence $J(a_0)$ for $a_0>0$. This maximum corresponds to $a_0=\mu$ and can be written in the form
\begin{equation}\label{Jc_res}
    J_c^\pm=
    \frac{\sqrt2LcH_{cm}\tau^{3/2}}{6\pi\sqrt3\lambda_0}
    \left(1\pm\Delta H\frac{L\sqrt{\tau}}{2\sqrt6H_{cm}\lambda_0}\right)
\end{equation}
Here the sign $+$ ($-$) corresponds to the current flowing parallel (anti-parallel) to the $y$ axis. 

The expression (\ref{Jc_res}) clearly shows the anisotropy of the critical current which differs for the two opposite directions of the current flow. The difference between the critical currents $J_c^\pm$ is proportional to the spontaneous magnetic field $\Delta H$ generated at the S/F interface due to the SOC. Note that the critical current is higher if the exchange field in the F layer is parallel to the magnetic field generated by the external current at $x=L$ and lower in the opposite case.
The equation~(\ref{Jc_res}) can be straightforwardly generalized for the case of arbitrary direction of the external current in the plane of the S/F structure. In this case the $\pm$ sign in the brackets should be replaced with $\cos\theta$ where $\theta$ is the angle between the direction of the current and the $y$ axis.

The predicted diode effect provides the alternative way for the experimental observation of the spontaneous currents generated by the SOC. To protect the S/F bilayer from the distraction caused by the heating effects one may use the pulse currents \cite{Bremer}.

	
\section{Conclusion}
\label{Concl}
To sum up, we develop the theory of superconducting phase transition in superconductor/ferromagnet bilayer with Rasba-type spin-orbit interaction at the S/F interface and $L \ll \xi$ (see Fig.\ref{Fig:sample}) using the Ginzburg-Landau approach. In the case of  $\lambda \ll L$ the phase transition is of the first order one even in the absence of external magnetic field. Moreover, its critical temperature is higher, than in bulk superconductor. In the external magnetic field ${\bf H}_0$  the critical temperature depends on mutual orientation of ${\bf H}_0$ and the exchange field inside the ferromagnet and the sign of spin-orbit coupling parameter: for positive (negative) SOC parameter it is higher (lower) for the parallel orientation in comparison with antiparallel one (see Fig.\ref{Fig:Tc_H0}). Both these results are manifestations of the spontaneous supercurrents flowing at the S/F interface \cite{Mironov} and can serve  hallmarks of these currents. Moreover, the dependence of the critical temperature on the magnetic configuration can be used for the experimental detection of the sign of SOC parameter. In the case of type-II superconducting layer the phase transition is the second order one, its critical temperature also increases in the absence of external magnetic field and depends on the relative orientation between external magnetic field and the exchange field, if the former is present. We also show that the critical current of the S/F bilayer reveals anisotropy in the plane of the layers.  The resulting diode-like effect may provide an alternative way for the experimental observation of the spontaneous superconducting currents generated by the SOC at the S/F interface.

\acknowledgements

The authors thank A. S. Mel'nikov and I. V. Zagorodnev for useful discussions. S.M. acknowledges the funding from Russian Science Foundation (Grant No. 20-12-00053, in part related to the calculations of the critical current). Zh.D. acknowledges the funding from Russian Science Foundation (Grant No. 18-72-10118, in part related to the calculations of the critical temperature). A.I.B. acknowledges support by the Ministry of Science and Higher Education of the Russian Federation within the framework of state funding for the creation and development of World-Class Research Center “Digital biodesign and
personalized healthcare” N075-15-2020-92. This work was supported by the French ANR OPTOFLUXONICS, EU COST CA16218 Nanocohybri, Foundation for the Advancement of Theoretical Physics and Mathematics BASIS (grants No. 18-1-3-58-1 and 18-1-4-24-1) and Russian Presidential Scholarship (Grants No SP-3938.2018.5 and No. SP-5551.2021.5).

\end{document}